\documentclass[pre,aps,a4paper,showpacs,superscriptaddress,floatfix,twocolumn]{revtex4-1}

\usepackage{amsmath}
\usepackage{amsfonts}
\usepackage{amssymb}
\usepackage{braket}
\usepackage{graphicx}
\usepackage{dcolumn}
\usepackage{bm}
\usepackage{textcomp}
\begin{document}

\def\Zz{\mathbb{Z}}
\def\T{\hat{T}}

\title{Modality of Equilibration in Non-equilibrium Systems}
\author{P. Barat }
\email{pbarat@vecc.gov.in}
\author{A. Giri}
\author{Nilangshu K. Das}
\author{ M. Bhattacharya}
\affiliation{Variable Energy Cyclotron Centre,1/AF Bidhannagar, Kolkata 700064, India}
\author{A. Dutta}
\affiliation{S.N. Bose National Centre for Basic Sciences, Block JD, Salt Lake, Kolkata 700098, India}
\date{\today }

\begin{abstract}
An open question in the field of non-equilibrium statistical physics  is whether there exists a unique way through which non-equilibrium systems equilibrate irrespective of how far they are away from equilibrium. To answer this question we have generated non-equilibrium states of various types of systems by molecular dynamics simulation technique. We have used a statistical method called system identification technique to understand the dynamical process of equilibration in reduced dimensional space. In this paper, we have tried to establish that the process of equilibration is unique.
\end{abstract}

\maketitle

For equilibrium systems, the probability distribution function (PDF) of energy is the well known Boltzmann distribution. Dynamic evolution of an equilibrium system through a sequence of equilibrium states is infinitesimally slow as it evolves in a quasistatic manner retaining the Boltzmann distribution at every stage. Practical systems are usually in the non-equilibrium state and try to reach its equilibrium quickly. The Boltzmann distribution is very robust and any non-equilibrium system left of its own tries to reach this distribution. Thermodynamic entities of equilibrium systems are studied by equilibrium statistical physics from the knowledge of the interaction of the microscopic constituents of the system. Small deviation of the equilibrium state by external perturbation in the linear response regime is analytically handled through equilibrium correlation function\cite{Onsager,Green,Kubo}. No generalized theoretical framework exists to calculate macroscopic thermodynamic properties of a system when it is far from equilibrium. In the last two decades, exact relations have been established relating thermodynamic parameters for systems irrespective of how far they are driven out of equilibrium. These studies are related to the path probabilities of the individual constituents of the non-equilibrium system\cite{Seifert}. Collective behavior of the constituents of non-equilibrium systems is poorly understood. Hence, it is important to study the evolution dynamics of non-equilibrium systems globally.  Probably this can be achieved by studying the time evolution of the PDF of energy of non-equilibrium systems as they approach to equilibrium.

It is easy to bring a system to its non-equilibrium state experimentally by the application of external perturbations like magnetic field, electric field, heat flux etc. However it is almost impossible to estimate the PDF of energy of the non-equilibrium system and its variation with time experimentally as one has to measure the energy of all the atoms in the system at different time. On the other hand it is quite easy to study the evolution of the PDF of energy for a non-equilibrium system and its approach to equilibrium by molecular dynamics (MD) simulation technique. In MD simulation, it is possible to calculate all the attributes of an individual atom at a finer time steps. Hence, we feel to study the modality of the equilibration process globally of a non-equilibrium state of a system, MD simulation is possibly the only way. In this work we report the mechanism of equilibration of the PDF of energy of several prototype systems by system identification technique when they are brought to non-equilibrium state using MD simulation and allowed to equilibrate of its own without any interaction with bath or sink. 
  
In MD simulation technique the dynamics of an individual atom is governed by the Newton's laws of motion. The interaction potentials among the atoms are calculated by various techniques from which the interaction force field is calculated. Newton's equations of motion are then integrated based on these derived force field to understand the dynamics of individual atoms. We have utilized the MD++\cite{MD++} simulation package for our MD simulation calculation. The simulations have been performed for elements having different crystal structures like Silicon (Si), Germanium (Ge) having diamond cubic structure, Iron (Fe), Molybdenum (Mo) having body centered cubic structure and solid Argon (Ar), Copper (Cu) and Aluminum (Al) having face centered cubic structure. Various interaction potentials for different elements have been used for the calculation. In each simulation we have adopted periodic boundary condition in three directions (x, y, z) in the simulation cell. Initially all the systems are equilibrated in the NVE (constant number of particles, volume and energy) ensemble at 100 K for 2.5 picoseconds (ps) and for solid Ar the system was equilibrated at 30 K for 10 ps. The attainment of equilibrium was obtained when the average kinetic energy (KE) is equal to the average potential energy (PE) of the atoms in the simulation cell.
\begin{table*}[!ht]
\caption{ The parameters used in the MD simulations.}
\label{table1}
\begin{center}
\begin{tabular}{lcccccccccc}
\hline
\hline
\\
\bf{Material} & \bf{Al} & \bf{Cu} & \bf{Fe} & \bf{Solid Ar} & \bf{Si-500} & \bf{Ge-500}& \bf{Si-800} & \bf{Ge-800}&\bf{Mo}\\\\
\hline
\\
Simulation cell size (in unit cell)  & $20^3$ & $20^3$  & $25^3$ &$20^3$ &$25^3$ & $15^3$& $15^3$& $15^3$ & $25^3$&\\\\
Number of atoms in simulation cell & 32000&32000 &31250 &32000 & 125000& 27000& 27000& 27000&31250&\\\\
Simulation time step (fs) &0.5 & 0.5&0.5 &2.0& 0.25&0.5&0.5&0.5 &0.5&\\\\
Inter-atomic potential used &GLUE &EAM\footnote{ EAM (Embedded Atom Model)} & FS\footnote{FS (Finnis-Sinclair)} &LJ\footnote{LJ (Lennard-Jones)} &SW\footnote{SW (Stillinger-Weber)} &SW&SW&SW &FS\\\\
Total number of data taken &3000 &3000 & 3000&10000 & 6000& 3000& 3000& 3000&3000&\\\\
No. of bins(n) used to define PDF of KE  &225 &200 & 200&130 & 325& 325& 325& 325&200&\\\\
Bin size(in Kelvin) &20 &20 & 20&5 & 20& 20& 20& 20&20&\\\\
\hline
\hline
\end{tabular}
\end{center}
\end{table*}

\begin{figure}[h!]
\centering
\includegraphics[width= 0.46 \textwidth]{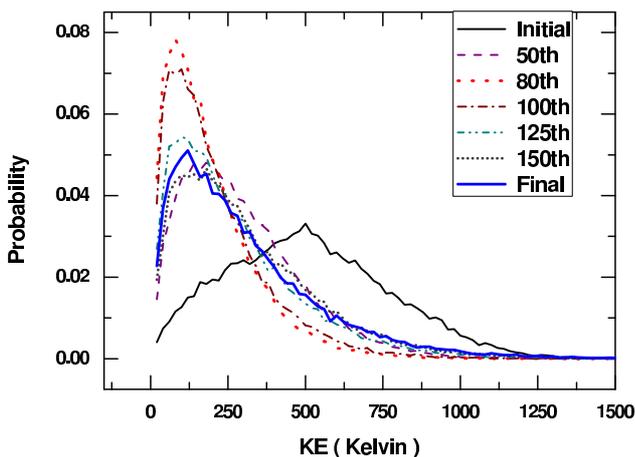}
\vspace{-10pt}
\caption{The probability distribution function of the kinetic energy (expressed in terms of temperature) of Cu atoms at initial, final and at five different time steps.}
\vspace{-5pt}
\label{figg1}
\end{figure}

To generate the non-equilibrium state the velocities of the individual atoms in the simulation cell are changed to random values maintaining  the average temperature at 500 K (70 K for solid Ar). The typical initial distribution function of KE for the Cu atoms in the non-equilibrium state is shown in Fig.~\ref{figg1}. The atoms in the simulation cell are then allowed to equilibrate. The KE of all the atoms are calculated in subsequent time steps to evaluate the PDF.  The PDF of KE for some discrete time steps are shown in Fig.~\ref{figg1}. After sufficient period of time the system is observed to attain an equilibrium temperature of 300 K (50 K for solid Ar). The corresponding PDF is also shown in Fig.~\ref{figg1}. To understand the effect of the average temperature of the non-equilibrium state, velocity of the atoms in Si and Ge cases are also randomised making the average temperature to 800K. The details of the parameters used in the MD simulations\cite{Pbarat} are given in the Table~\ref{table1}.

\begin{figure}[h!]
\begin{center}
\includegraphics[width= 0.46\textwidth]{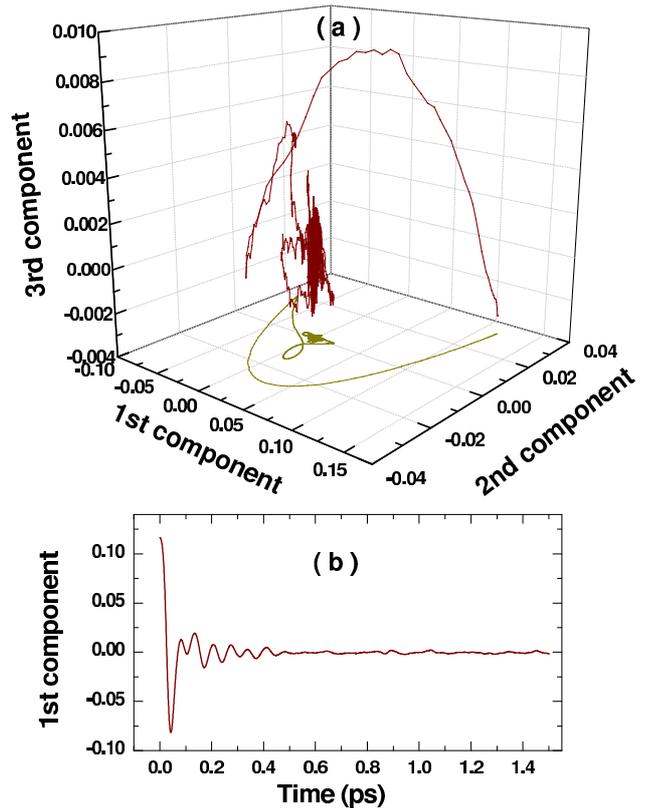}
\vspace{-10pt}
\caption{(a) Time evolution of the PDF of kinetic energy in the subspace spanned by the three eigen functions corresponding to the three largest eigen values of the covariance matrix ${\varSigma}_x $. The curve in the bottom plane is its projection. (b) Time evolution of the primary principal component of Cu.}
\vspace{-5pt}
\label{figg3}
\end{center}
\end{figure}

 The variation of the KE of an atom is dictated by the force field experienced by it. The primary force field experiences by an atom arises from the nearest neighbor interaction potential whereas the interaction arising from the distant atoms generates a random force field and there is a frictional force field which makes the atom to lose its KE. Thus the dynamics of the individual atom can be represented by the Langevin type equation\cite{Hasegawa}. Because of the existance of these force fields the KE of an atom and its reflection to the PDF of the ensemble at any instant of time will certainly have a deterministic and a random components. The calculated PDF of KE at different instant of time from the MD simulation is in the form of the distribution of KE in the bins and is identified as a vector $\Ket {X_i}$ of dimension n in the $i^{th}$ time step. The covariance matrix ${\varSigma}_x=\frac{1}{N} {\sum}_{i=1}^N \Ket {X_i}\Bra {X_i}-\Ket {\bar{X}}\Bra {\bar{X}} $ of all such vectors for a particular element is calculated.
Where, 
$\Ket {\bar{X}}= \frac{1}{N}{\sum}_{i=1}^N\Ket {X_i}$
  is the mean vector and N is the total number of time steps required for equilibration. Thus, each component of the data vector $\Ket {X_i}$  consists of a deterministic component $\Ket {V_i}$  and a random noise component $\Ket {\xi_i}$. As $\Ket {\xi_i}$ is uncorrelated with $\Ket {V_i}$ and with another component of the random noise, we can write the covariance matrix as
${\varSigma}_x={\bar{\varSigma}}_x+\langle {\xi}^2\rangle I_{n}$.
 Where, ${\bar{\varSigma}}_x $ is the covariance matrix formed by the deterministic components $\Ket {V_i}$  and $I_{n}$ is the unit matrix of dimension n. Thus $\lambda_j$, the eigenvalue of ${\varSigma}_x$, will be related with the corresponding eigenvalue $ {\bar{\lambda}}_j$ of  ${\bar{\varSigma}}_x $ by the relation
$\lambda_j={\bar{\lambda}}_j+\langle {\xi}^2\rangle $,
where j=1 to n. Hence, the noise causes all the eigenvalues of the covariance matrix ${\varSigma}_x$ to be non-zero. Thus, when the data vectors are reconstructed with the eigenfunctions corresponding to the eigenvalues of significant magnitude they are devoid of random noise.
\begin{figure}[h!]
\includegraphics[width= 0.46\textwidth]{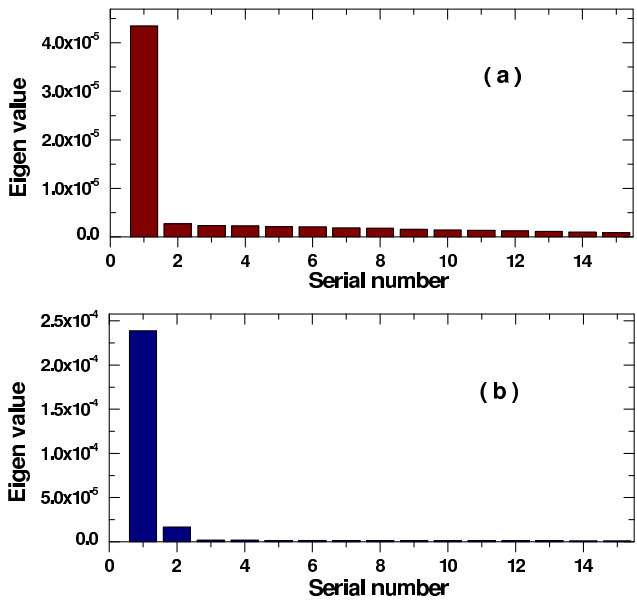}
\vspace{-10pt}
\caption{Eigen value spectra of the covariance matrix of (a) Ar and (b) Cu.}
\vspace{-5pt}
\label{figg2}
\end{figure}
 Typical eigen values or the principal component spectra of ${\varSigma}_x $, for solid Ar and Cu are shown in Fig.~\ref{figg2}. The prominent principal components for all the cases studied are found to be only two or three except for solid Ar where it is more than 20(Fig.~\ref{figg2}). This is because, solid Ar being an inert gas crystal, the deterministic force field is reasonably weak compared to that of other studied elements. The trajectory of the evolution of the PDF in the subspace spanned by the eigen-functions corresponding to the primary principal components and the variation of the first principal component with time for Cu is shown in Fig.~\ref{figg3}. The time evolution of these principal components for all the crystalline elements have a distinct oscillatory signature and a typical representation is shown in Fig.~\ref{figg3}b. 

During the process of equilibration the PDF of KE oscillates about the equilibrium Boltzmann distribution as shown in Fig.~\ref{figg1}. The amplitude of this oscillation reduces with time as the system approaches towards the equilibrium. As the Boltzmann distribution is obtained on the concept of the most probable distribution by a combinatorial way, any system left of its own will certainly approach this distribution irrespective of the nature of the interaction among its constituents. Thus the dynamics of equilibration in our studies is ultimately related to the nature of the coupled oscillation of the bins of PDF around their respective mean positions, determined by the equilibrium Boltzmann distribution. The existence of only 2 or 3 prominent eigen values of the covariance matrix ${\varSigma}_x$ suggests that the oscillations are highly correlated. This modality of transforming the dynamics in reduced dimensional space helped us to analyze the complex dynamics in a more elegant way. The oscillation of the most significant eigen value (containing 82\% to 95\% of the trace of ${\varSigma}_x$ in the respective cases) shows a decayed oscillation (Fig.~\ref{figg3}b) as the system approaches equilibrium. This suggests that at equilibrium the deterministic component  of PDF goes to zero. However, we have seen that for all studied cases the noise component remains the same. The observed nature of this oscillation is caused by the concerned system. So the prime interest will be to identify the nature of the system that is responsible for this oscillation by system identification technique\cite{Ljung}. This technique uses statistical methods to generate mathematical models of dynamical systems linking the observed data from the system.

In the MD simulation we have purposely disturbed the equilibrium distribution of KE of the ensemble of atoms at time $t = 0$. Therefore the time evolution of the PDF can be considered as an outcome of the finite impulse response to the KE distribution in the systems. Response of a system to a finite impulse can be used to estimate the characteristics of the system concerned.  From these observed impulse responses we identify the systems responsible for this manifestation. In system identification technique, the impulse response of a dynamical  system is modeled from the response data and the system is estimated non-parametrically by time domain correlation analysis\cite{Ljung}. Correlation analysis presupposes a linear system and it does not require any explicit model structure.
\begin{table*}[ht!]
\caption{ The frequencies at the poles of the transfer functions for different elements and their corresponding Debye frequencies at 298K.}
\label{table2}
\begin{center}
\begin{tabular}{l|ccccccccc}
\hline
\hline
\\
 & \bf{Al} & \bf{Cu} & \bf{Fe} & \bf{Solid Ar} & \bf{Si-500} & \bf{Ge-500}& \bf{Si-800} & \bf{Ge-800}&\bf{Mo}\\\\
\hline
\\

\bf{1st frequency} (\bf{THz}) & $7.13$ &	$6.68$   &	$8.11$    &	$1.74$    &	$11.59$    &	$8.11$ &  $11.33$  &  $8.34$   &  $9.96$ \\\\
\bf{2nd frequency} (\bf{THz})& $12.03$  &	$14.80$    &	$14.20$    &	$4.19$    &	$30.67$    &	$14.35$ &  $29.11$  &  $14.48$  &  $15.91$\\\\
\bf{3rd frequency} (\bf{THz}) & $17.21$  &	   &	   &	   &	$33.60$    &	$18.96$ &  $33.60$  &  $18.84$  &   \\\\
\bf{Debye frequency}& $8.13$ 	& $6.46$  & $7.76$  &		$1.92$ \footnote{At 0K}\cite{Stewart}   & $14.4$  &  $8.4$ &	   &	 &	$7.86$ \\
\bf{at 298K} (\bf{THz})\cite{Ho} &\\

\hline
\hline
\end{tabular}
\end{center}
\end{table*}
 The response to a finite impulse $u(t)$ is equal to the convolution of the impulse response and the transfer function $h(t)$ of the system and is given as
$ y(t)=\int_{\tau=0}^{+\infty}h(\tau)u(t-\tau) \,d\tau $.

\begin{figure}[h!]
\begin{center}
\includegraphics[width= 0.43\textwidth]{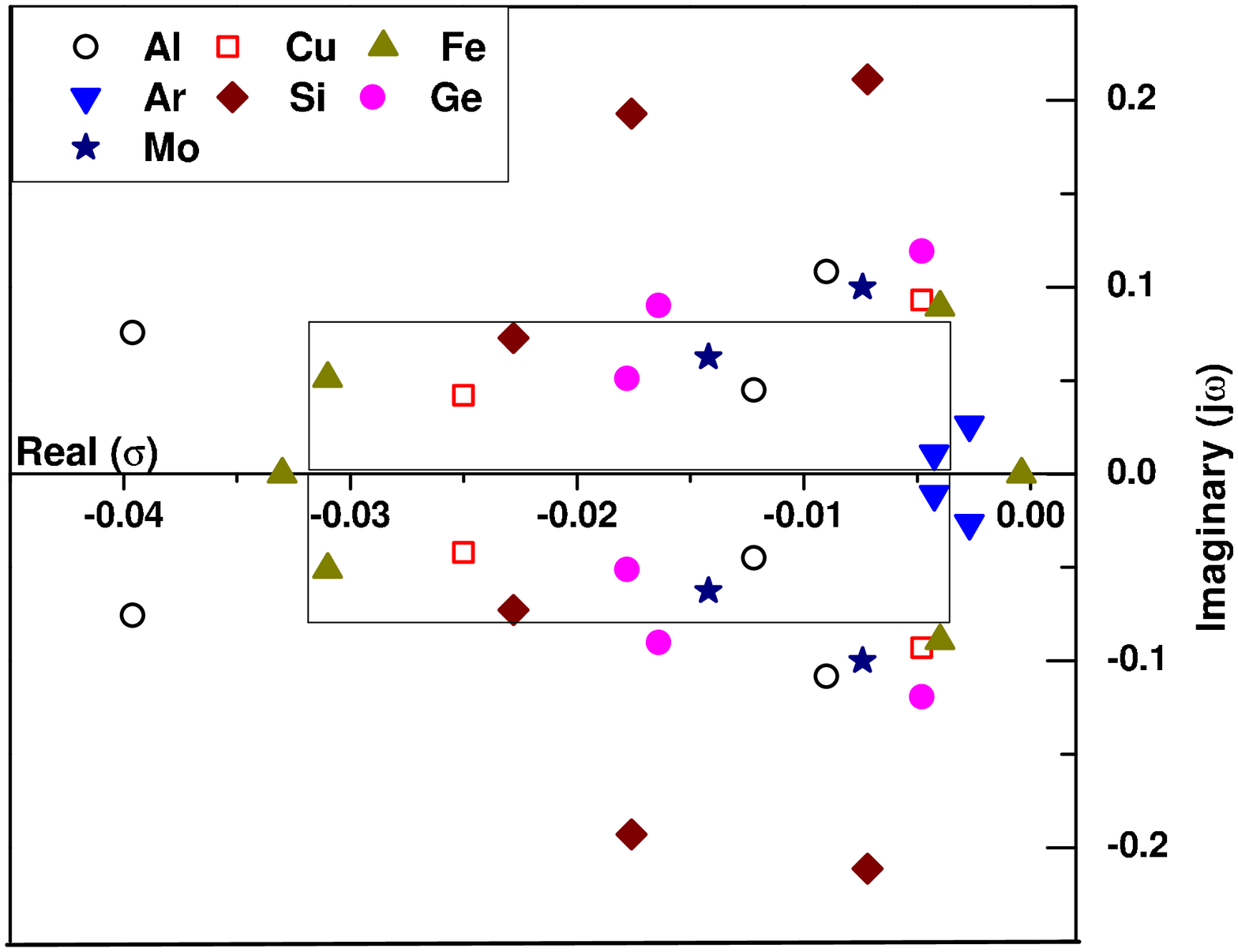}

\vspace{-10pt}
\caption{Real and imaginary part of the poles of the transfer functions for different elements. The points inside the box represents lower frequency. The poles of Si-800 and Ge-800 are not shown as they are almost identical to those of Si-500 and Ge-500 respectively.}
\vspace{-5pt}
\label{figg4}
\end{center}
\end{figure}

  We have identified the transfer function of the system in Laplace domain by $H(s)= \beta \frac{{\Pi}_{i}(s-z_i)}{{\Pi}_{j}(s-p_j)}$, where $\beta$ is the gain constant. $H(s)$ is a rational function of the complex variable $s=\sigma+j\omega$. It provides the response characteristics of the system in the continuous domain without solving the necessary differential equation, governing the dynamics. The $z_i$'s are the zeros and $p_j$'s are the poles of $H(s)$, since $H(p_j)=\infty$ provided $z_i \ne p_j$. The poles and zeros of $H(s)$ together with $\beta$ provide a complete description of the system. The partial fraction extension of $H(s)$ can be written as $H(s)= \sum_{j=1}^PK_j\frac{1}{(s-p_j)}$. Where, $K_j$'s are the residues of the particular pole $p_j$ and P is the number of poles in $H(s)$. From the inverse Laplace transformation of $H(s)$ one can obtain the natural response of the system as $\sum_{j=1}^PK_j\exp(p_jt)$, where the poles $p_j$'s are the natural frequencies\cite{Sedra} of the system. The expression of $H(s)$ for Cu is given as
\begin{widetext}
 \[H(s)=0.0022\frac{(s+1.018)(s+0.0296)((s+0.0078)^2+0.0842^2)}{(s+0.3178)(s+0.1044)((s+0.0048)^2+0.0930^2)((s+0.0250)^2+0.0420^2)} \]
\end{widetext}
 For our studied systems, the transfer functions have 2 to 3 complex conjugate pole pairs that are located to the left plane of the $j\omega$ axis as shown in Fig.~\ref{figg4} and the corresponding frequencies are given in Table~\ref{table2}. This guarantees that the denominator of $H(s)$ is never zero for any non-negetive $\sigma$. As a consequence the system can not generate a sustained sine wave oscillation of the principal component but it will always be a decaying wave as shown in Fig.~\ref{figg3}b. The poles of solid Ar are nearer to the origin directing the system to equilibrate over a longer period of time. This is the manifestation of the weak interaction of the atoms of the inert gas system. The poles and the corresponding frequencies of $H(s)$ for Si-500 and Si-800 and for Ge-500 and Ge-800 are almost identical[Table~\ref{table2}], although their initial non-equilibrium states are different.

The system in the process of equilibration redistributes its KE among the atoms. This in turn changes the spatial configuration of the atoms and the PE of the system. The impetus of the oscillation of PDF arises because the system is being compelled to attain the Boltzmann distribution to equalise the average PE equal with the average KE. The lowest frequencies of the poles of $H(s)$ for different elements are very near to the Debye frequencies\cite{Ho,Stewart}. Other observed frequencies are much higher compared to the frequency of the lattice vibration. The lowest frequency having higher $\sigma$ value signifies that it attenuates faster and helps the process of equilibration to the maximum extent. At the frequencies of the poles, the transfer function blows up and then only energy can be exchanged freely back and forth between KE and PE.

Equilibrium state of a system is uniquely defined. Non-equilibrium state of a system can be generated by various means and it is possible to bring it to any state far away from its equilibrium state. Thus, it is not realistic to adopt a proper definition for a non-equilibrium system. Consequently a question can arise whether there exists a unique way through which a non-equilibrium system equilibrates irrespective of how far it is away from equilibrium. In this paper we have tried to answer this question. For this reason we have not only generated non-equilibrium states of various elements but also generated different non-equilibrium states for the same element and allowed it to equilibrate. The ensemble taken, is the simplest of all i.e. NVE or the micro-canonical ensemble. Our statistical analysis of the time evolution of the PDF of KE shows that all the systems equilibrate in a unique way irrespective of the nature of the system and how far it is away from equilibrium.



\begin{thebibliography} {99}
\bibitem{Onsager}L. Onsager, Phys. Rev. \textbf{37}, 405 (1931); \textbf{38},  2265 (1931).
\bibitem{Green}M. S. Green, J. Chem. Phys. \textbf{19}, 1036 (1951)
\bibitem{Kubo}R. Kubo, J. Phys. Soc. Jpn. \textbf{12}, 570 (1957)
\bibitem{Seifert} Udo Seifert, Rep. Prog. Phys. \textbf{75}, 126001 (2012) and the references therein.
\bibitem{MD++} http//micro.stanford.edu
\bibitem{Pbarat}P. Barat, A. Giri, M. Bhattacharya, Nilangshu K. Das, and A. Dutta, Europhys. Lett. \textbf{104}, 50003 (2013).
\bibitem{Hasegawa}Hideo Hasegawa, Phys. Rev. E \textbf{83}, 021104 (2011) and the references therein.
\bibitem{Ljung} Lennart Ljung, {\it System Identification: Theory for the User} (second edition), Prentice Hall PTR (1999).
\bibitem{Sedra}Adel S. Sedra and Kenneth Carless Smith, {\it Microelectronic Circuits}, 4th ed, New York:Oxford University Press, 1998
\bibitem{Ho} C. Y. Ho, R. W. Powell, and P. E. Liley, Journal of Physical and Chemical Reference Data, Volume \textbf{3},  Supplement no. 1 (1975).
\bibitem{Stewart}G. R. Stewart, Rev. Sci. Instrum. \textbf{54},  1 (1983).
\end{thebibliography}
\end{document}